# The Role of Dwell Time on Advancing 3D Nano-printing of Complex Iron Nanostructure Geometries using Focused Electron Beam Induced Deposition

Sameh Okasha*, and Trevor P. Almeida

School of Physics and Astronomy, University of Glasgow, Glasgow, G12 8QQ, United Kingdom
∗Corresponding authors: sameh.okasha@glasgow.ac.uk, trevor.almeida@glasgow.ac.uk

The shift from two- to three-dimensional structures is not only a prominent trend in nanomagnetism but also reflects a broader movement across nanotechnology as a whole [1,2]. Fields such as nanoelectronics [3], nanophotonics [4], data storage are poised to benefit from a new generation of greener, more versatile and multifunctional technologies enabled by this transition to 3D structures. While significant challenges remain, recent progress in bottom-up lithography, advanced microscopy, and computational techniques has made the future realization of these advancements increasingly feasible. Focused electron beam-induced deposition (FEBID), as shown in figure 1, is a cutting-edge direct-write nanofabrication technique used to manipulate matter at the nanoscale [5–8], combined with the unique magnetic properties of nanomaterials. It has gained significant attention in recent decades due to its potential applications in magnetic memories, such as racetrack memory. Consequently, substantial efforts have been directed toward developing nanofabrication techniques and characterizing magnetic nanoelements of various metal depositions. The high vapor pressure of precursors for materials such as Pt [1] and Co [9] is particularly advantageous for 3D nanoprinting under electron-limited conditions [7,10], where a high precursor flux is beneficial. Consequently, 3D-FEBID performs best with precursors that exhibit high vapor pressure [1]. In contrast, Fe deposition presents challenges due to its precursors having relatively lower vapor pressure. One reported solution was increasing the pressure; however, it has the opposite effect, hindering 3D nanoprinting [11].

This necessitates refining the deposition parameters to overcome these challenges, opening the door to accurate studies of deposition rates as shown in figure 2. In our work, Fe-based nanostructures have been fabricated precisely using iron pentacarbonyl ($Fe(CO)_5$) as a precursor. These structures include nanowires, 5-µm bridges, rings, tetrapods, spirals and flower-like nanostructures. The deposition parameters, including electron beam dwell time, beam current, and precursor flux, have been further refined, have been further refined, enabling the precise fabrication of complex geometries.

The growth calibrations, morphology, and composition of iron nanostructures produced by FEBID are influenced by factors such as electron beam current, precursor flow conditions, and the balance between reaction-rate-limited and adsorption-dominant processes. Our work examines the deposition environment and its optimization to address challenges associated with the low yield and growth rate of Fe, including factors such as precursor temperature, substrate temperature, resolution, magnification, and surface conductivity. The optimization of electron beam parameters involves refining conditions such as voltage, current, focus, and deposition angle. As shown in Figure 2, calibration of the deposition process allowed us to determine key parameters of the deposition model equation = $G_r e^{-KR} \exp\left(\frac{-d^2}{2\sigma^2}\right) t$, where Gr is the growth rate, K is the thermal resistance scaling factor, and σ is the standard deviation of the deposit width at distance, d. These parameters were extracted by analyzing experimental deposition profiles and comparing them with simulation results to refine the predictive accuracy of the model. Several SEM images, accompanied by analytical explanations, were utilized to determine the initial values of these model parameters. We also introduce effective simulated dwell-time graphs and highlight their role in advancing the 3D nanoprinting of complex Fe geometries by refining the model parameters to more precise values as shown in figure 3. This refinement enables the deposition of more intricate and accurate Fe structures. The deposited structures were further analysed using aberration-corrected TEM to determine their chemical composition through EDX, ED, and EELS.
The results of this work pave the way for further studies on the magnetic properties of 3D magnetic nanostructures, which could serve as the foundation for a new class of non-volatile memory storage devices. Fe nanostructures created through FEBID have applications in diverse fields, including magnetic data storage devices, magnetic force microscopy tips, sensors, and fundamental research areas such as magnetic frustration phenomena, curvilinear magnetism, magnonics, and fluxonics [12,13].

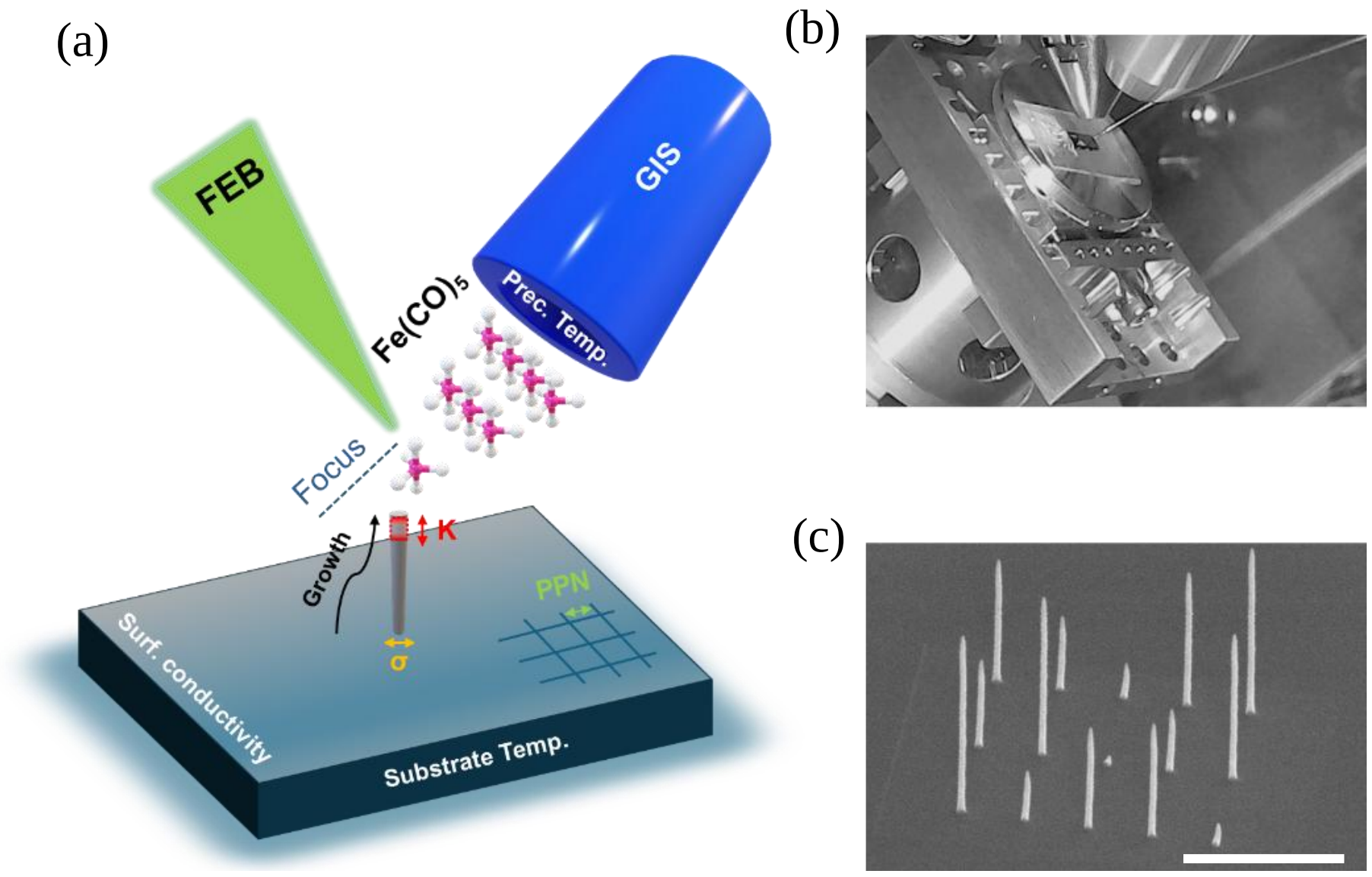

**Fig. 1.** Schematic illustration of the FEBID process. (a) Calibration procedure, detailing the determination of model parameters such as electron beam settings and precursor dynamics. (b) Image of varying deposition angles which influences the growth morphology. (c) SEM image of an array of nanowires used for calibration and optimization. Scale bar = 1 um

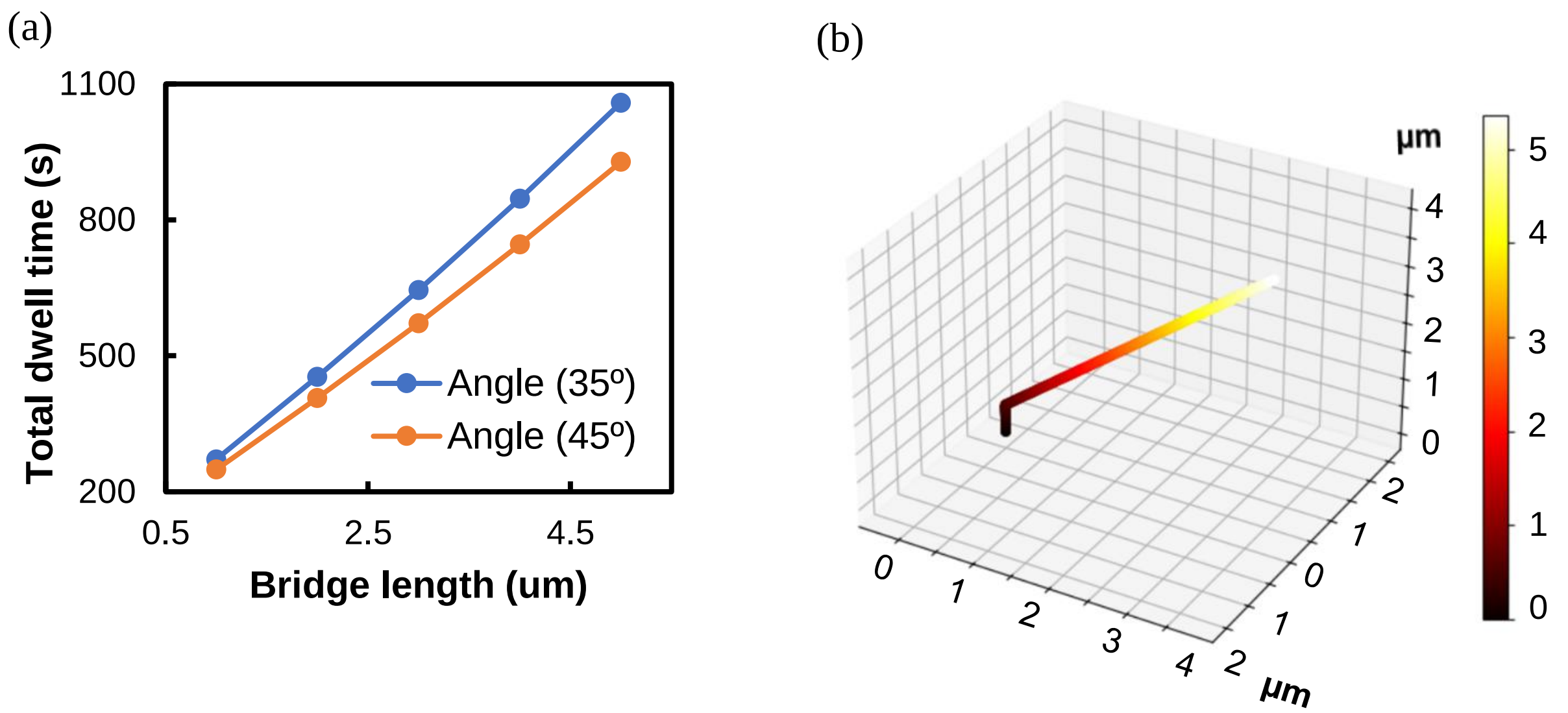

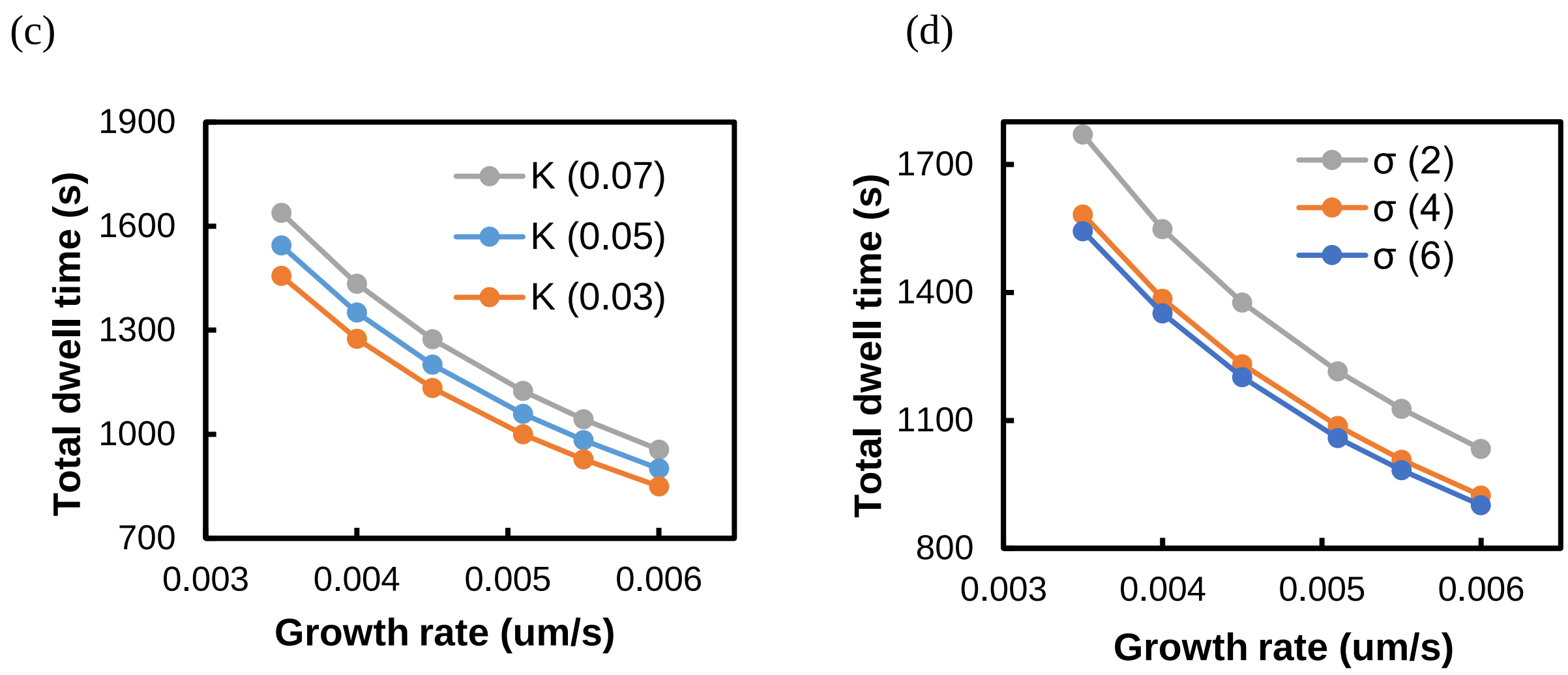

**Fig. 2.** The role of dwell time on advancing 3D nano printing of Fe 5 um bridge using FEBID. (a) Relationship between deposition angle and nanowire length, (b) the heat resistance simulation model, (c) thermal resistance scaling factor (K), (d) the standard deviation of the deposit width (σ).

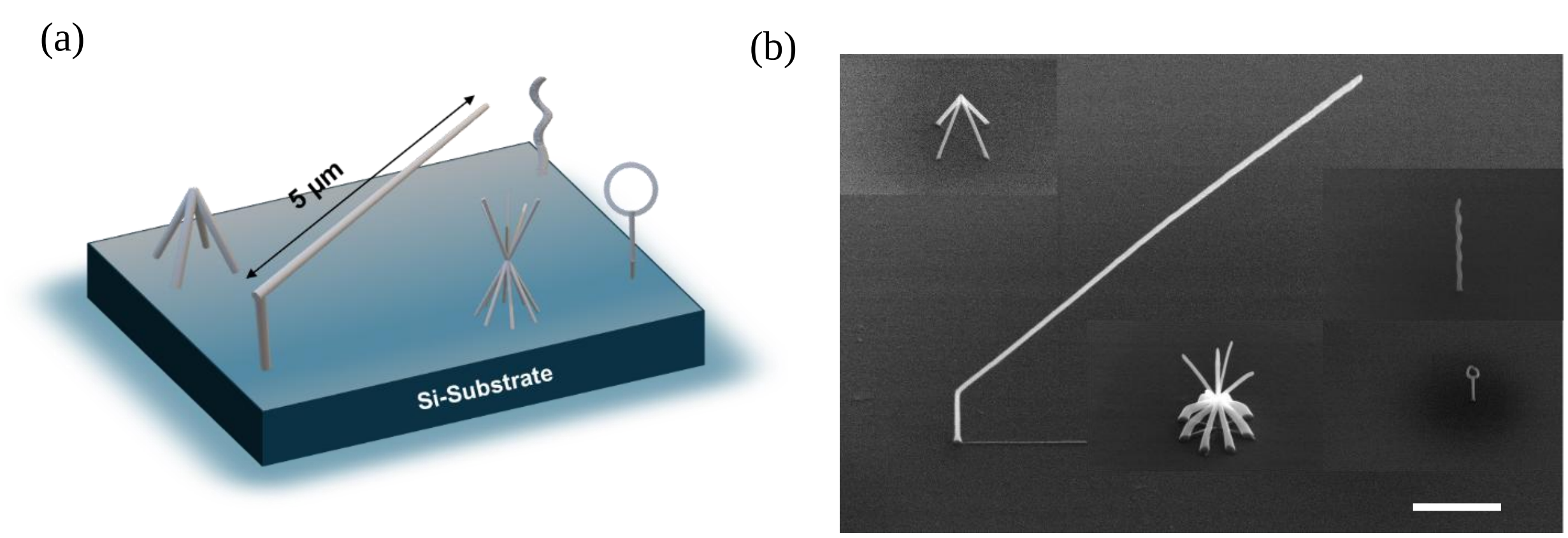

**Fig. 3.** Results of 3D nano-printing using FEBID. (a) A 3D model of the structure designed using OpenSCAD software. (b) SEM images of the fabricated 3D iron structures, showing the achieved morphology. Scale bar = 1 μm.

14. The authors acknowledge funding from the Engineering and Physical Science Research Council (Grant ref: EP/X025632/1)